\documentclass[10pt]{article}
\usepackage{graphicx,amsmath,amssymb,amsbsy,latexsym,amsthm}
\usepackage{psfrag}

\theoremstyle{plain}
\newtheorem{theorem}{(analogue of) Theorem}
\newtheorem{lemma}{(analogue of) Lemma}
\newtheorem{corollary}{Corollary}

\newcommand{\startproof}{\setlength{\parindent}{0in}\textbf{Proof.} }
\newcommand{\finishproof}{\hfill $\blacksquare$ \\}

\def\R{\mathbb{R}}

\def\C{\mathbb{C}}
\def\N{\mathbb{N}}

\def\tr{\mathrm{tr}}
\def\half{\frac{1}{2}}

\newcommand{\lalg}[1]{\mathfrak{#1}}  
\newcommand{\so}{\mathfrak{so}}

\newcommand{\eqa}{\begin{eqnarray}}
\newcommand{\neqa}{\end{eqnarray}}
\newcommand{\be}{\begin{equation}}
\newcommand{\ee}{\end{equation}}

\newcommand{\Ref}[1]{(\ref{#1})}

\newcommand{\scrM}{\mathcal{M}}

\newcommand{\scrB}{\mathcal{B}}

\newcommand{\scrZ}{\mathcal{Z}}

\newcommand{\dual}{\,\,{}^\star\!}

\newcommand{\dif}{\mathrm{d}}

\newcommand{\double}[2]{\hspace{0.1em} #1 \hspace{#2} #1 \hspace{0.1em}}
\newcommand{\ident}{\double{1}{-0.35em}}

\newcommand{\Hil}{\mathcal{H}}

\newcommand{\Link}{\mathrm{Link}}

\newcommand{\node}{n}
\newcommand{\intlor}{N}

\begin{document}

\title{Regularization and finiteness of the Lorentzian LQG vertices}
\author{Jonathan Engle${}^{ab}$ and Roberto Pereira${}^b$
 \\[1mm]
\normalsize \em ${}^a$ MPI f\"{u}r Gravitationsphysik (Albert-Einstein-Institut), Am M\"{u}hlenberg 1, D-14476 Potsdam, EU \\
\normalsize \em ${}^b$ CPT%
\footnote{Unit\'e mixte de recherche (UMR 6207) du CNRS et des Universit\'es
de Provence (Aix-Marseille I), de la Meditarran\'ee (Aix-Marseille II) et du Sud (Toulon-Var); laboratoire affili\'e \`a la FRUMAM (FR 2291).} , CNRS Case 907, Universit\'e de la M\'editerran\'ee, F-13288 Marseille, EU
}
\maketitle

\begin{abstract}
We give an explicit form for the Lorentzian vertices
recently introduced for possibly defining the dynamics of
loop quantum gravity.  As a result of
so doing, a natural regularization of the vertices is suggested.  The
regularized vertices are then proven to be finite.  An interpretation of the
regularization in terms of a gauge-fixing is also given.
\end{abstract}

\section{Introduction}

In the search for quantum gravity, loop quantum gravity \cite{lqgrevs}
has provided a well-understood kinematical framework, arising from standard
quantization methods, with the assumption that parallel transports
have well-defined operator analogues in the quantum theory.  Spin-foams
have been proposed as an approach to the dynamics of the theory that
retains manifest space-time covariance (for reviews and some useful original
papers, see \cite{sfrevs, baez}). Through works of the past year
\cite{ls_coh, eprlett, eprpap, fk, ls_model, ep, pereira, elpr},
progress was made in modifying the more traditional Barrett-Crane
model \cite{bc_eucl, bc_lorentz}, by addressing the issue of the simplicity constraints with
more care.  As a result, the kinematics of the models \cite{eprlett, eprpap, pereira, elpr},
covering all values of the Immirzi parameter in both Euclidean and Lorentzian signatures,
exactly matched those of loop quantum gravity.\footnote{
This is not true for all the models proposed in the above works. See
\cite{fk, ep}.}

However, the issue of the finiteness of the
Lorentzian LQG spin-foam vertices was not addressed in the papers
\cite{pereira, elpr}.
We address the issue in this present paper.
In this paper we show that Lorentzian LQG
spin-foam vertices possess an overall multiplicative factor equal to the volume of $SL(2,\C)$,
so that without regularization, the vertices are indeed infinite.
However, in computing expectation values of quantities, these volume factors
will just cancel.  Therefore, one can simply drop the overall volume
factor in the vertex.
Such a regularization can be independently justified via a gauge-fixing interpretation,
similar to but different from that in \cite{fl2004}; this is discussed in appendix \ref{gfapp}.
%
%
%
We prove that with this regularization,
the vertex is finite. In both proposing the regularization and proving finiteness,
an explicit form of the LQG vertices derived in section \ref{reviewsect} is key. This new form has formal similarities to the Barrett-Crane Lorentzian vertex, which allows some of
the reasoning of \cite{bb} to be used also for proving the finiteness of the LQG vertices.
Adaptations of the relevant arguments from \cite{bb} are summarized in the form of a lemma
and two theorems in appendix \ref{proofs}.  The rest of the proof of finiteness of the LQG vertices is then presented in section \ref{maintextproof} in the main text. 

For both the Euclidean and Lorentzian LQG models,
there still remains the issue of finiteness of the full state-sum for a fixed
triangulation. We leave this issue for future work. We would like to remark that the analysis presented here does not directly apply to the Lorentzian model proposed in \cite{fk};
nevertheless it could provide a foundation for analysis of the Lorentzian model in \cite{fk}.

The paper is organized as follows.  We start with a brief review of the Lorentzian LQG
spin-foam models for finite $\gamma$, starting from a triangulation with space-like tetrahedra;
in so doing, a more concrete and explicit approach is taken than that originally taken in \cite{elpr}.
This then aides in suggesting the regularization, proposed in the subsequent section. In the last section we prove
finiteness, in part reusing reasoning from the paper \cite{bb}. Finally, we close with a summary
and brief discussion.  In appendix \ref{gfapp} is presented the gauge-fixing interpretation of the regularization.

\section{Review of the models}
\label{reviewsect}

Here we review the classical discrete theory, canonical quantization
of the constraints, and path integral dynamics from \cite{elpr}, deriving a
new expression for the vertex in the process.

\noindent\textit{Classical discrete theory}

The starting point is a Regge-like discretization of first
order gravitational variables.  The continuum first order
variables one considers are those of Plebanski theory \cite{plebanski,baez}:
an $SL(2,\C)$ connection $\omega$ and an $\lalg{sl}(2,\C)$-valued
2-form $B$ on space-time.  This two form is then subjected to so-called simplicity
constraints \cite{plebanski,baez} which ensure that it is of the form
$B = \dual e \wedge e$ for some co-tetrad $e$.

To construct the Regge-like discretization, introduce a
triangulation $\Delta$ of space-time $\scrM$
by (oriented) 4-simplices. We will denote typical
4-simplicies, tetrahedra, and triangles in $\Delta$
respectively by $v, t, f$. The basic variables
then consist in an $SL(2,\C)$ group element $V_{vt}(\equiv V_{tv}^{-1})$
for each 4-simplex $v$ and tetrahedron $t$ therein,
and an $\lalg{sl}(2,\C)$ algebra element $B_f(t)$ for
each tetrehedron $t$ and triangle $f$ therein.
$V_{vt}$ is to be heuristically understood as the
parallel transport map, determined by $\omega$, from the
tetrahedron $t$ to the center of the 4-simplex $v$.
$B_f(t)$ is to be understood as the integral of $B = \dual e \wedge e$
on the triangle $f$, in the frame at $t$.  (For details,
see \cite{eprpap, ep, elpr}.)

It is convenient to furthermore define, for each triangle $f$ and each
pair of tetrahedra $t,t' \in \Link(f)$,
\begin{displaymath}
U_f(t,t') := V_{tv_1} V_{v_1 t_1} V_{t_1 v_2} \cdots V_{v_n t'}
\end{displaymath}
where the product is around the link in the clock-wise direction
from $t'$ to $t$. The constraints on the variables are then
\begin{enumerate}
\item $U_f(t,t') B_f(t') = B_f(t) U_f(t,t')$ \qquad $\forall$ $f$ and
$t,t' \in \Link(f)$
\item {(discrete simplicity constraint)}\\
$\exists$ an assignment of a timelike $n^I_t$ to each $t$, such that
\begin{equation}
\label{discsimp}
C_{ft}^I := n_{tJ}\left(\dual B_f\right)^{JI} \approx 0 \quad \forall f \in t .
\end{equation}
\end{enumerate}
Note that the simplicity constraint (\ref{discsimp}) implies
\textit{both} of the more traditional simplicity constraints,
the so-called diagonal simplicity constraint
$\tr\left[(\dual B_f(t))B_f(t)\right]
\approx 0$ and cross-simplicity constraints
$\tr\left[(\dual B_f(t))B_{f'}(t)\right]
\approx 0 \quad \forall f,f' \in t$.\footnote{
Thanks to Laurent Freidel for pointing out this fact
about the classical theory.
}
In the prior literature
\cite{eprlett,eprpap,elpr}, the fact that (\ref{discsimp})
implies the diagonal simplicity constraint as well the
cross-simplicity constraint was overlooked, so that diagonal
simplicity was imposed separately.
We will see that these
observations extend to the
quantum theory as well: the quantum version of (\ref{discsimp}),
appropriately understood, will be seen to already contain within it the
quantum diagonal simplicity constraint! Additionally,
as noted in \cite{elpr}, (\ref{discsimp}) is able to distinguish
between the $B = \pm \dual e \wedge e$ and $B = \pm e \wedge e$
sectors of Plebanski theory, selecting only the first
%
%
of these sectors.

The above constraints are incorporated as follows:
(1.) is imposed prior to varying the action, while
(2.) is first solved canonically and then the result inserted in the path integral
\footnote{The
closure constraint, $\sum_{f\in t} B_f(t) = 0$, which is dealt
with more directly in other presentations, is exactly recovered only at the quantum level,
by the integration over connection variables in the path-integral dynamics.}.

Next consider a 3-surface $\Sigma$ consisting of
tetrahedra in the triangulation $\Delta$; call this
triangulation $\Delta_3$.  Let $\gamma_{\Sigma}$ denote
the graph dual to $\Delta_3$.  We will denote typical
links and nodes in $\gamma_{\Sigma}$ by $\ell, n$, respectively.
The canonical phase space $\Gamma_{\Sigma}$ associated with
$\Sigma$ is then labelled by the basic variables
$B_\ell(n) \in \so(3,1)$, $U_\ell(n,n') \in SL(2,\C)$.
%
%
Define array of $\lalg{sl}(2,\C)$
matrices $\tau^{IJ} = -\tau^{JI}$ by
\begin{eqnarray}
\nonumber
\tau^{i0} &=& \frac{1}{2}\sigma^i \\
\tau^{ij} &=& \frac{-i}{2}\epsilon^{ij}{}_k \sigma^k
\end{eqnarray}
Define $J_\ell(\node) = \frac{1}{16\pi G}\left( B_\ell({\node})
+ \frac{1}{\gamma}\dual
B_\ell({\node})\right)$. The non-zero Poisson brackets are then given by
\begin{eqnarray}
{\{J_\ell({\node}')^{IJ}, U_\ell({\node},{\node}')\}} &=& U_\ell({\node},{\node}') \tau^{IJ} \\
{\{J_\ell({\node})^{IJ}, U_\ell({\node},{\node}')\}} &=& \tau^{IJ} U_\ell({\node},{\node}') \\
{\{J_\ell({\node})^{IJ}, J_\ell({\node})^{KL}\}} &=& \lambda^{[IJ][KL]}{}_{[MN]} J_\ell(n)^{MN}
\end{eqnarray}
where $\lambda^{[IJ][KL]}{}_{[MN]}$ denotes the structure constants in the basis $\tau^{IJ}$.


\noindent\textit{Quantization}

The quantization of $\Gamma_{\Sigma}$ leads us to the
kinematical space of states
\begin{equation}
\Hil_{\Sigma} = L^2\left(SL(2,\C)^{|L(\gamma_\Sigma)|}\right) .
\end{equation}
Let $\hat{J}_\ell({\node})^{IJ}$ denote the
right-invariant vector fields, determined by the
basis $\tau^{IJ}$ of $\mathfrak{sl}(2,\C)$, on the copy of
$SL(2,\C)$ associated with the link $\ell$, with orientation such that the node
${\node}$ is the source of $\ell$. The $B_\ell(n)$'s are then
represented by
\begin{equation}
\hat{B}_\ell({\node}) := 16 \pi G
\left(\frac{\gamma^2}{\gamma^2+1}\right)
\left(\hat{J}_\ell({\node})-\frac{1}{\gamma}\dual \hat{J}_\ell({\node})\right)
\end{equation}
As in \cite{elpr}, to solve the simplicity constraint, we gauge-fix
the normal $n_t^I$ in (\ref{discsimp}) to be $n_t^I \equiv n^I := (1,0,0,0)$.
The simplicity constraints \Ref{discsimp} are then imposed by appropriately
quantizing the `master constraint'
\begin{equation}
\label{clmaster}
M_{n\ell} := \sum_{i=1}^3 \left[(\dual B_f(t))^{0i}\right]^2
\end{equation}
associated to each node $n$ and incident link $\ell$.
To quantize and solve this, it is convenient to
introduce a basis of $\Hil_{\Sigma}$ adapted to the constraint. First,
we recall that if $\Hil_{N, \rho}$ is the carrying space for
a Lorentz group irrep $(N, \rho)$ in the principal series, one can
decompose $\Hil_{N, \rho}$ into irreps of the $SU(2)$ subgroup preserving $n^I$,
arriving at
\begin{equation}
\label{candecomp}
\Hil_{N, \rho} = \oplus_{k \ge N/2} \Hil_k
\end{equation}
where $\Hil_k$ is the carrying space for the spin $k$ $SU(2)$ irrep
appearing in the decomposition.  Using this, we construct
a basis of generalized $SL(2,\C)$ spin-networks; specifically,
these will be the projected spin-networks of \cite{projsn} with
the normal gauge-fixed to be $n^I$. Given an assignment of
a Lorentz irrep $(\intlor_\ell, \rho_\ell)$ in the principal series to
each link, an $SU(2)$ spin $k_{n\ell}$ for each specification of a node and incident link,
and an $SU(2)$ intertwiner $i_{n}$ among the four $SU(2)$ irreps $\{k_{n\ell}\}_{\ell \in n}$
at $n$, we define
\begin{eqnarray}
\nonumber
\Psi_{\{{\intlor}_\ell,\rho_\ell;k_{n\ell},i_{n}\}}(U_\ell)
&\equiv&
\langle U_\ell \mid \{ {\intlor}_\ell, \rho_\ell; k_{n\ell}, i_{n} \} \rangle \\
\label{projspindef}
&:=& \left(\bigotimes_\ell   D^{({\intlor}_\ell, \rho_\ell)}(U_\ell) \cdot \bigotimes_{n}
\left[\left(\otimes_{\ell\in n} P_{k_{n\ell}}\right)\otimes i_{n}\right]\right).
\end{eqnarray}
where $P_k$ is the projector onto the spin-$k$ component in the
decomposition (\ref{candecomp}).  Note in this expression that at each node
$\left[\left(\otimes_{\ell\in n} P_{k_{n\ell}}\right)\otimes i_n\right]$ is a tensor in
$\otimes_{\ell\in n} \Hil_{({\intlor}_\ell, \rho_\ell)}$; the role of the
labels $\{k_{\ell n}\}_{\ell \in n}$ and $i_n$ at each node is to
specify a tensor among the four Lorentz irreps on the adjacent edges.
Contracting these all together gives the desired
generalized $SL(2,\C)$ spin-network (\ref{projspindef}).
Note, in particular, that even though $SL(2,\C)$ representations
in the principal series are infinite dimensional, the incorporation
of the projection operators $P_k$ in (\ref{candecomp}) ensures that all contractions
involve effectively only finite sums, so that the right hand side of
(\ref{projspindef}) is guaranteed to be finite.
Finally, let $\hat{L}^i_{\ell n} := \half \epsilon^i{}_{jk} \hat{J}_\ell^{jk}(n)$.
The basis (\ref{projspindef}) then diagonalizes the operators
\begin{eqnarray*}
\left(\hat{J}_\ell\cdot \hat{J}_\ell\right) \Psi_{\{{\intlor}_\ell,\rho_\ell;k_{n\ell},i_n\}} &=& \half({\intlor}_\ell^2 - \rho_\ell^2 -4)\Psi_{\{{\intlor}_\ell,\rho_\ell;k_{n\ell},i_n\}}\\
\left(\hat{J}_\ell\cdot \dual\hat{J}_\ell\right) \Psi_{\{{\intlor}_\ell,\rho_\ell; k_{n\ell},i_{n}\}} &=& {\intlor}_\ell \rho_\ell \Psi_{\{{\intlor}_\ell,\rho_\ell; k_{n\ell}, i_n\}}\\
\hat{L}_{n\ell}^2 \Psi_{\{{\intlor}_\ell,\rho_\ell;k_{n\ell},i_n\}} &=& k_{n\ell}(k_{n\ell}+1) \Psi_{\{{\intlor}_\ell,\rho_\ell;k_{n\ell},i_n\}}
\end{eqnarray*}
In terms of this basis, the master constraint \Ref{clmaster}, quantized as
in \cite{elpr}, is
\begin{equation}
\hat{M}_{n \ell} \Psi_{\{{\intlor}_\ell,\rho_\ell;k_{n\ell},i_{n}\}} =
\left[\left(1+\frac{1}{\gamma^2}\right)\left(k_{n\ell}^2 -(N_\ell/2)^2\right)
+\frac{1}{4\gamma^2}\left(\rho_\ell - \gamma N_\ell\right)^2\right]
\Psi_{\{{\intlor}_\ell,\rho_\ell;k_{n\ell},i_{n}\}}
\end{equation}
As $k_{n\ell} \ge N_{\ell}/2$, solving this constraint forces both
of the terms on the right hand side to separately vanish. In all,
simplicity thus implies
\begin{equation}
\label{finalsimp}
k_{\ell_- \ell} = \frac{N_\ell}{2} = \frac{\rho_\ell}{2\gamma} = k_{\ell_+ \ell}
\end{equation}
for all $\ell$, and where $\ell_-, \ell_+$ denotes respectively the
source and target of $\ell$. Because $k_{n\ell}$ is the quantum number for the non-Lorentz
scalar quantity $\hat{L}^2_{n\ell}$, this is not an $SL(2,\C)$ invariant equation.
This lack of $SL(2,\C)$ invariance derives from the gauge-fixing of $n^I$ and will be relevant
in appendix \ref{gfapp}.

\noindent\textit{Path integral dynamics}

Consider the case when $\Delta$ consists in a single
4-simplex, and let $\gamma$ denote its boundary graph.
The vertex amplitude is derived as the amplitude for
a generalized spin-network state on the boundary.
We begin by writing down the amplitude for
BF theory, reflecting the flatness equation of motion present
in BF theory:
\begin{equation}
A[U_{tt'}] = \int d V_{vt} \Pi_{tt'} \delta(U_{tt'}V_{t'v} V_{vt}).
\end{equation}
multiplying by a generalized spin-network (\ref{projspindef}) and integrating
over the $U_{tt'}$ with the Haar measure leads to the amplitude
for a generalized spin-network on the boundary of a single 4-simplex $v$
\begin{eqnarray}
\nonumber
A_v[{\intlor}_f,\rho_f;k_{tf},i_t] &=&
\int d V_{vt} \Psi_{\{{\intlor}_f, \rho_f; k_{tf}, i_t\}}^{\text{4-simplex}}
(V_{tv}V_{vt'}) \\
\nonumber
&=&
\left(P^{\text{gauge}}_{SL(2,\C)}\Psi_{\{{\intlor}_f, \rho_f; k_{tf}, i_t\}}^{\text{4-simplex}}
\right)\big|_{\text{Triv. Conn.}} \\
\label{bfvertex}
&\equiv& 15j_{\scriptscriptstyle SL(2,\C)}({\intlor}_f, \rho_f;
I^t(k_{tf}, i_t))
\end{eqnarray}
where $I^t(k_{tf}, i_t) := P^t_{SL(2,\C)}
\left[\left(\otimes_{f \in t} P_{k_{ft}}\right)\otimes i_t \right]$,
and where $P^t_{SL(2,\C)}$ denotes a formal group averaging
over $SL(2,\C)$ gauge transformations at $t$.
At each $t$, $I^t(k_{tf}, i_t)$ is thus
formally an $SL(2,\C)$ intertwiner.

Combining with the simplicity constraints, we obtain an
$SU(2)$ LQG spin-foam model with partition function
\begin{equation}
Z_{\rm GR}= \sum_{j\!{}_{f}, i_t}\ \prod_f (2j_f)^2 (1+\gamma^2)
\prod_v A_v(j\!{}_{f}, i_t)
\label{Z}
\end{equation}
with $j\!{}_{f} \in \N/2$ and
\begin{equation}
\label{lqgvertex}
A_v(j\!{}_{f},i_t) := A_v[2j_f, 2\gamma j_f; j_f, i_t] =
15j_{\scriptscriptstyle SL(2,\C)}(2j_f, 2\gamma j_f;
I^t(j_f, i_t))
\end{equation}
This is the vertex amplitude for general $\gamma$, as in \cite{elpr}.
Setting $\gamma = 0$ gives the flipped Lorentzian model \cite{pereira}.

\section{Regularization}

Is the vertex (\ref{lqgvertex}) and/or (\ref{bfvertex}) finite?  The answer is no.
However, it is not hard to see that an observation similar to that in
\cite{bc_lorentz} can be used to regularize it: the vertex consists in
an integral over five copies of the group, but one of these is redundant.
That is, if
we perform any four of the five integrals, the result is independent
of the fifth integration variable, so that the last integral is
redundant.\footnote{This comes from a $SL(2,\C)$ gauge invariance acting at each vertex, see appendix \ref{gfapp}.}

To demonstrate, number the tetrahedra $1,2,3,4,5$, and label the
5 group integration variables as $V_1, V_2, V_3, V_4, V_5$.
Dropping the fifth integration, we symbolically write
\begin{equation}
\label{vertexreg}
A_v^{\text{reg}}[{\intlor}_f,\rho_f;k_{tf},i_t] :=
\int d V_1 d V_2 d V_3 d V_4
\begin{minipage}{1.5in}
\psfrag{a45}[c][][.55][0]{$V_4^{-1}V_5$}
\psfrag{a34}[c][][.55][0]{$V_3^{-1}V_4$}
\psfrag{a51}[c][][.55][0]{$V_5^{-1}V_1$}
\psfrag{a41}[c][][.55][0]{$V_4^{-1}V_1$}
\psfrag{a35}[c][][.55][0]{$V_3^{-1}V_5$}
\psfrag{a52}[c][][.55][0]{$V_5^{-1}V_2$}
\psfrag{a12}[c][][.55][0]{$V_1^{-1}V_2$}
\psfrag{a24}[c][][.55][0]{$V_2^{-1}V_4$}
\psfrag{a13}[c][][.55][0]{$V_1^{-1}V_3$}
\psfrag{a23}[c][][.55][0]{$V_2^{-1}V_3$}
\includegraphics[height=1.5in]{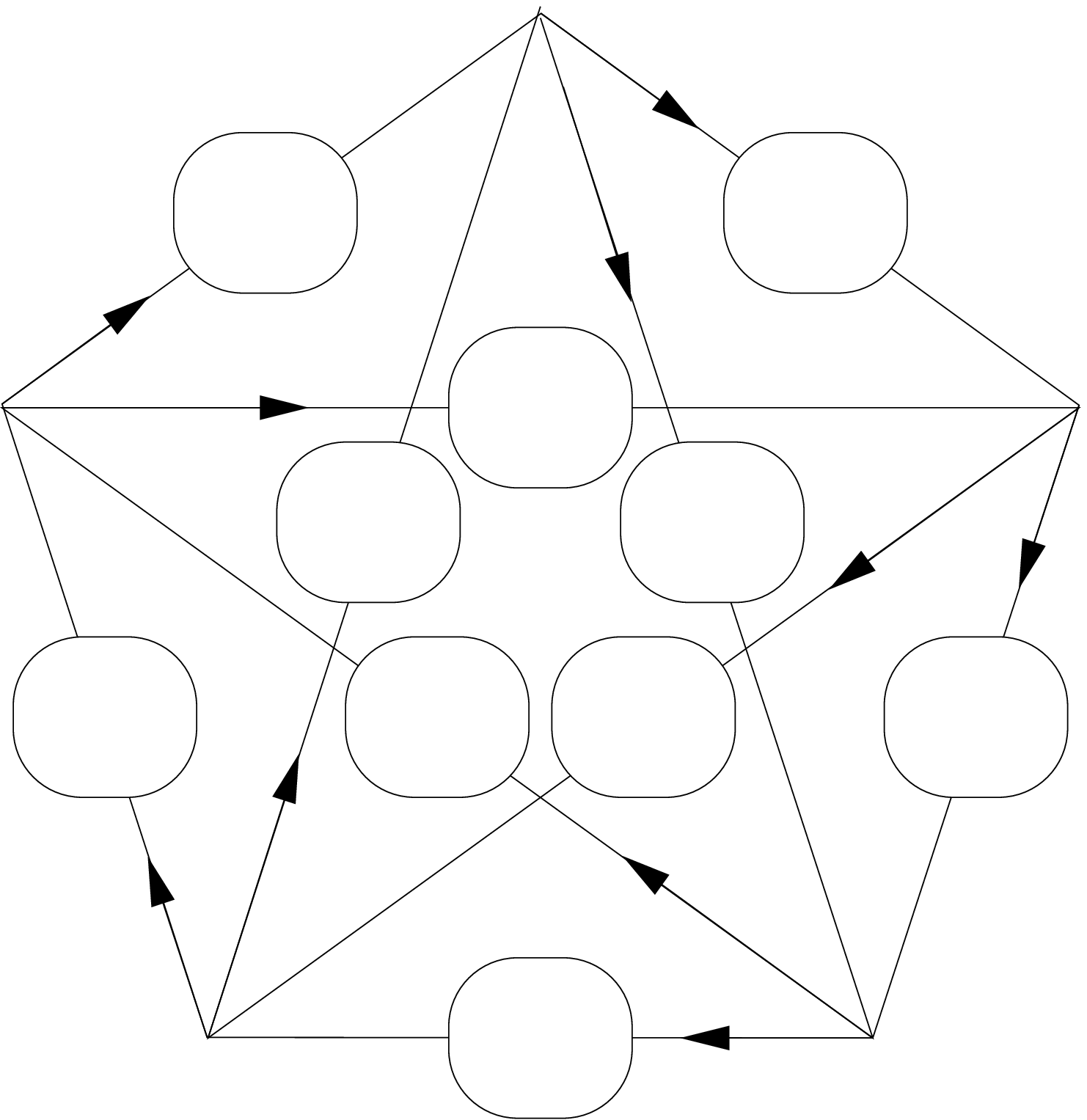}
\end{minipage}
\end{equation}
Performing the change of variables
\begin{eqnarray}
V_1 &\mapsto \tilde{V}_1 =& V_5^{-1} V_1 \\
V_2 &\mapsto \tilde{V}_2 =& V_5^{-1} V_2 \\
V_3 &\mapsto \tilde{V}_3 =& V_5^{-1} V_3 \\
V_4 &\mapsto \tilde{V}_4 =& V_5^{-1} V_4,
\end{eqnarray}
using the right invariance of the Haar measures,
and noting that for $i,j=1,..,4$,
$\tilde{V}_i^{-1} \tilde{V}_j = V_i^{-1} V_j$,
the expression (\ref{vertexreg}) simplifies to
\begin{equation}
\label{simpreg}
A_v^{\text{reg}}[{\intlor}_f,\rho_f;k_{tf},i_t] =
\int d V_1 d V_2 d V_3 d V_4
\begin{minipage}{1.5in}
\psfrag{a45}[c][][.55][0]{$V_4^{-1}$}
\psfrag{a34}[c][][.55][0]{$V_3^{-1}V_4$}
\psfrag{a51}[c][][.55][0]{$V_1$}
\psfrag{a41}[c][][.55][0]{$V_4^{-1}V_1$}
\psfrag{a35}[c][][.55][0]{$V_3^{-1}$}
\psfrag{a52}[c][][.55][0]{$V_2$}
\psfrag{a12}[c][][.55][0]{$V_1^{-1}V_2$}
\psfrag{a24}[c][][.55][0]{$V_2^{-1}V_4$}
\psfrag{a13}[c][][.55][0]{$V_1^{-1}V_3$}
\psfrag{a23}[c][][.55][0]{$V_2^{-1}V_3$}
\includegraphics[height=1.5in]{integrandfig1.eps}
\end{minipage},
\end{equation}
which is manifestly independent of the
unintegrated group element $V_5$.
Thus, the last integral, when performed, simply introduces a factor
equal to the volume of $SL(2,C)$, which is infinite. We regularize it by simply dropping the last integral\footnote{which, depending on how the $V$'s are numbered,
could be any one of the five integrals.  The result is independent
of which one you drop.}. Incorporating the simplicity
constraints in the form \Ref{finalsimp}, we thus propose
\begin{equation}
A_v^{\text{reg}}(j_f, i_t) := A_v^{\text{reg}}[2j_f,2\gamma j_f;j_f,i_t] .
\end{equation}

\section{Proof of finiteness}
\label{maintextproof}

For each set of labels $\{\rho_f, N_f; k_{tf}, i_t\}$ determining
a projected spin-network on the 4-simplex boundary graph,
let $\tilde{F}_{\rho_f, N_f; k_{tf}, i_t}: SL(2,\C)^5 \rightarrow \C$ denote
the integrand in the definition of the associated vertex \Ref{vertexreg}.
That is, define
\begin{equation}
\label{Ftildedef}
\tilde{F}_{\rho_f, N_f; k_{tf}, i_t}(V_1, V_2, V_3, V_4, V_5)
:=
\begin{minipage}{1.5in}
\psfrag{a45}[c][][.55][0]{$V_4^{-1}V_5$}
\psfrag{a34}[c][][.55][0]{$V_3^{-1}V_4$}
\psfrag{a51}[c][][.55][0]{$V_5^{-1}V_1$}
\psfrag{a41}[c][][.55][0]{$V_4^{-1}V_1$}
\psfrag{a35}[c][][.55][0]{$V_3^{-1}V_5$}
\psfrag{a52}[c][][.55][0]{$V_5^{-1}V_2$}
\psfrag{a12}[c][][.55][0]{$V_1^{-1}V_2$}
\psfrag{a24}[c][][.55][0]{$V_2^{-1}V_4$}
\psfrag{a13}[c][][.55][0]{$V_1^{-1}V_3$}
\psfrag{a23}[c][][.55][0]{$V_2^{-1}V_3$}
\includegraphics[height=1.5in]{integrandfig1.eps}
\end{minipage}
\end{equation}
where the schematic on the right hand side represents the projected
spin-network with its $10$ $SL(2,\C)$ arguments. Next, every element
$V \in SL(2,\C)$ can be decomposed
\begin{equation}
\label{sl2cdecomp}
V = B(x) R
\end{equation}
for some $R \in SU(2)$ and some boost $B(x)$.  We here parametrize
the boosts by a point $x$ in the
hyperboloid $H$ of future directed unit time-like vectors in Minkowski space;
$B(x)$ denoting the unique boost mapping $e:=(1,0,0,0)$ to $x$.
Decomposing each of the
arguments of $\tilde{F}$ in the manner \Ref{sl2cdecomp},
and using the $SU(2)$ invariance of the projected spin-network
on the right hand side of (\ref{Ftildedef}) to drop the rotations,
\begin{equation}
\tilde{F}_{\rho_f, N_f; k_{tf}, i_t}(V_1, \dots V_5)
= \tilde{F}_{\rho_f, N_f; k_{tf}, i_t}(B(x_1), \dots B(x_5)) .
\end{equation}
Next, for each $\rho, N, k\ge N/2$ and $m_{\pm} \in \{-k, -k+1, \dots, k\}$, define
\begin{equation}
\label{Kdef}
K^{\rho, N}_{k m,k'm'}(x_1, x_2)
:= D^{\rho,N}_{k m,k'm'}(B(x_1)^{-1}B(x_2)) .
\end{equation}
%
%
Let us label faces $f$ in $v$ by the two tetrahedra $(tt')$ they bound,
so that each $f$ is labeled by an unordered pair of numbers $1,2,3,4,5$.
%
%
Furthermore, let us label pairs $(t,f) \in v$ by \textit{ordered} pairs of
tetrahedra $(t,(tt'))=:tt'$, and hence ordered pairs of numbers $1,2,3,4,5$.
For each assignment of $\rho_{(tt')}, N_{(tt')}$ to faces $f \in v$ and
$k_{tt'}, m_{tt'}$ to pairs $(t,f) \equiv (t,(tt')) \in v$, define
\begin{equation}
\label{Fdef}
\left[F_{\rho_{(tt')}, N_{(tt')}; k_{tt'}}\right]_{\{m_{tt'}\}} := \int \dif x_1 \dif x_2 \dif x_3 \dif x_4
\prod_{\substack{t,t' \in \{1,2,3,4,5\} \\ t<t'}}K^{\rho_{(tt')}, N_{(tt')}}_{k_{tt'}m_{tt'},k_{t't}m_{t't}}(x_{t}, x_{t'}) .
\end{equation}
where $\dif x$ denotes the volume form on the hyperboloid.  One can check that in terms of
the decomposition \Ref{sl2cdecomp}, the Haar measure decomposes as
\begin{equation}
\dif V = \dif R \dif x
\end{equation}
where $\dif R$ is the Haar measure on $SU(2)$.
Using this equation, one sees the vertex is equal to
\begin{equation}
\label{Fvertex}
A_v^{\text{reg}}[\rho_f, N_f; k_{ft}, i_t] = \left[\otimes_{t \in v} i_t\right] \cdot \left[F_{\rho_{f}, N_{f}; k_{ft}}\right].
\end{equation}
Because the contraction sums on the right hand side are all finite, it is sufficient to
prove finiteness of the elements $\left[F_{\rho_{f}, N_{f}; k_{ft}}\right]_{\{m_{(tf)}\}}$
%
%
in order to prove
finiteness of the vertex. We shall do this, using arguments very similar to \cite{bb}.

Let us now look at the boosts entering (\ref{Fdef}).  First we rewrite
the composition of two boosts $B(x_1)^{-1}$, $B(x_2)$
as
\begin{equation}
\label{boostcomp}
B(x_1)^{-1}B(x_2) = R(x_1,x_2) B_z(r(x_1, x_2)) R'(x_1, x_2)
\end{equation}
for some two rotations $R(x_1, x_2)$, $R(x_1, x_2)$ and boost $B_z(r(x_1,x_2))$
in the $z$-direction, where $r(x_1,x_2)$ denotes the rapidity of the boost.
%
%
We can always choose this decomposition such that $r(x_1,x_2)$ is positive,
and we do so.
%
%
$r(x_1,x_2)$ is in fact the hyperbolic distance between $x_1$ and $x_2$.
To see this, we recall that the hyperbolic distance, or hyperbolic angle, between
two points $x_1, x_2 \in H$ is defined by
\begin{equation}
d(x_1, x_2) := \cosh^{-1} (x_1, x_2)
\end{equation}
where $( \cdot, \cdot )$ denotes the Minkowski metric.  We thus have
\begin{eqnarray}
\nonumber
\cosh d(x_1, x_2) &=& (x_1, x_2) = (B(x_1)e, B(x_2)e) = (e, B(x_1)^{-1} B(x_2)e) \\
\nonumber
&=& (e, R(x_1,x_2) B_z(r(x_1,x_2)) R'(x_1, x_2) e) = (R(x_1, x_2)^{-1}e, B_z(r(x_1, x_2)) R'(x_1, x_2)e) \\
\nonumber
&=& (e, B_z(r(x_1,x_2)) e) \\
&=& \cosh r(x_1,x_2)
\end{eqnarray}
so that
%
%
$r(x_1, x_2) = d(x_1, x_2)$, proving $r(x_1, x_2)$ is
the hyperbolic distance, as claimed.

Now, let us consider the matrix elements of $B_z(r)$ in a given
representation $(\rho, N)$ in the principal series.  We use the
canonical basis for the carrying space, \textit{i.e.} the basis
diagonalizing $\hat{L}^2$ and $\hat{L}_z$: $\hat{L}^2 |k,m\rangle = k(k+1) |k,m\rangle$,
$\hat{L}_z |k,m\rangle = m |k,m\rangle$. Because the generator $K_z := J^{03}$ of
$z$-boosts commutes with $\hat{L}_z$, we have
\begin{equation}
\label{zboostmat}
D^{\rho,N}_{k m, k' m'}(B_z(r)) = \delta_{m m'} d^{\rho,N}_{k k' m}(r)
\end{equation}
for some function $d^{\rho, N}_{k k' m}(r)$.
As shown in appendix \ref{lorentz}, the behavior of $d^{\rho, N}_{k k' m}(r)$ in the $r \rightarrow \infty$ limit is of the form
\begin{equation}
\label{dasymp}
d^{\rho,N}_{k k' m}(r) \propto e^{- \lambda_{N,m} r}
\end{equation}
where
\begin{equation}
\lambda_{N,m} = 1 + \left|m + \frac{N}{2}\right| \ge 1.
\end{equation}
This in particular implies that for any $\epsilon > 0$,
\begin{equation}
\lim_{r\rightarrow \infty} e^{(1-\epsilon)r} d^{\rho,N}_{k k' m}(r) = 0.
\end{equation}
Because $e^{(1-\epsilon)r}d^{\rho,N}_{kk'm}(r)$
is furthermore continuous,
%
%
we know $e^{(1-\epsilon)r} d^{\rho,N}_{k k' m}(r)$
is bounded on $r \in [0,\infty)$, so that there exists $C^{\rho,N,\epsilon}_{kk' m} \in \R^+$
such that
\begin{eqnarray}
\nonumber
e^{(1-\epsilon)r} d^{\rho,N}_{k k' m}(r) &<& C^{\rho,N,\epsilon}_{kk' m} \\
\Rightarrow \qquad d^{\rho,N}_{k k' m}(r) &<& C^{\rho,N,\epsilon}_{kk'm} e^{-(1-\epsilon)r}
\end{eqnarray}
for all $r \in [0,\infty)$.
%
%

Consider next the matrix elements of the rotations in (\ref{boostcomp}).
From p.63 in \cite{ruhl}, for $R$ a rotation,
\begin{equation}
\label{rotmat}
D^{\rho, N}_{km,k'm'}(R) = \delta_{k k'} D^k_{m m'}(R)
\end{equation}
where $D^k_{m m'}(R)$ are the matrix elements in
the spin-$k$ representation of $SU(2)$.
Because matrix elements on the right hand side are in a \textit{unitary}
representation, and we are using an orthonormal basis, all of these matrix elements
have absolute value less than or equal to
one.\footnote{To
see that this is true for a general unitary matrix $U$,
and orthonormal basis $x_i$,
from $U U^\dagger = \ident$ we have $\langle x_i, U U^\dagger x_j\rangle = \delta_{ij}$,
so that for $i=j$ we have
\begin{eqnarray*}
1 &=& \langle x_i, U U^\dagger x_i \rangle = \sum_k \langle x_i, U x_k \rangle \langle x_k, U^\dagger x_i \rangle
= \sum_k \langle x_i, U x_k \rangle \overline{\langle x_i, U x_k \rangle} \\
&=& \sum_k \left| \langle x_i, U x_k \rangle \right|^2
\end{eqnarray*}
from which $\left| \langle x_i, U x_k \rangle \right|^2 \le 1$.}

Let us put the above observations together.
From (\ref{boostcomp}), (\ref{zboostmat}), (\ref{rotmat}), we have
\begin{eqnarray}
\label{comp_matel}
D^{\rho, N}_{k m, k' m'}\left(B(x_1)^{-1}B(x_2)\right)
= \sum_{m''\in \{-\text{min}\{k,k'\},-\text{min}\{k,k'\}+1, \dots, \text{min}\{k,k'\}\}}
D^k_{m m''}(R) d^{\rho,N}_{k k' m''}(r(x_1,x_2)) D^{k'}_{m'' m'}(R')
\end{eqnarray}
where the one sum has been made explicit. We then have
\begin{eqnarray}
\nonumber
\left| D^{\rho, N}_{k m, k' m'}\left(B(x_1)^{-1}B(x_2)\right)\right|
&\le& \sum_{m''}\left| D^k_{m m''}(R) d^{\rho,N}_{k k' m''}(r(x_1,x_2)) D^{k'}_{m'' m'}(R')\right| \\
\nonumber
&=&  \sum_{m''} \Big| D^k_{m m''}(R)\Big| \left| d^{\rho,N}_{k k' m''}(r(x_1,x_2))\right|
\left| D^{k'}_{m'' m'}(R')\right| \\
\nonumber
&\le& \sum_{m''} \left|d^{\rho,N}_{k k' m''}(r(x_1,x_2))\right| \\
\label{ineq1}
&<& \left(\sum_{m''} C^{\rho,N,\epsilon}_{kk'm''}\right) e^{-(1-\epsilon)r} .
\end{eqnarray}
Defining $C^{\rho,N,\epsilon}_{kk'} := \sum_{m}C^{\rho,N,\epsilon}_{kk'm}$,
which is finite because the sum is finite, we thus have
\begin{equation}
\label{Kbound}
\left|K^{\rho,N}_{kk',mm'}\right| = \left| D^{\rho, N}_{k m, k' m'}\left(B(x_1)^{-1}B(x_2)\right)\right|
< C^{\rho,N,\epsilon}_{kk'} e^{-(1-\epsilon)r}
\end{equation}
for all $r \in [0,\infty)$.
This bound \Ref{Kbound}, given the expression
\Ref{Fdef} for $\left[F_{\rho_f,N_f;k_{ft}}\right]_{m_{ft}}$,
allows us to adapt the arguments of Baez and Barrett in \cite{bb} to show that $\left[F_{\rho_f,N_f;k_{ft}}\right]_{m_{ft}}$ is finite.

Let us summarize how the arguments of Baez and Barrett can be used.
The bound \Ref{Kbound} is the analogue of Lemma 1 in \cite{bb}. Lemma 2,3 and 4
in \cite{bb} can be used again without change.  One can prove the analogue
of Lemma 5, and Theorems 2 and 3 in \cite{bb} using logic analogous to that
in \cite{bb}.  For completeness, we present these analogues
in appendix \ref{proofs}. The desired finiteness of
$\left[F_{\rho_f,N_f;k_{ft}}\right]_{m_{ft}}$ then comes as a corollary.
As was already noted, this in turn is then sufficient to prove that the vertex amplitude
\Ref{Fvertex} is finite for all labels $\{\rho_f, N_f, k_{ft}, i_t\}$ on the 4-simplex graph.
Note this finiteness of the vertex holds \textit{even prior} to imposing the simplicity
constraints \Ref{finalsimp}; nevertheless it is the case when \Ref{finalsimp} is satisfied
that ultimately concerns us.\footnote{The finiteness in the unconstrained
case may be useful for defining a Lorentzian BF-theory model; but in a BF-theory model,
one will be summing over all possible $k_{tf}$ in the state-sum, which is an infinite sum
for each pair $(tf)$.} Note this proves finiteness of the vertex for all finite gamma, as
well as for the flipped case.


\section{Discussion}

In this paper, by writing the Lorentzian vertex of \cite{elpr} in a more
concrete manner, we were able to see a natural way to regularize the vertex.
We then proved the vertex, so regularized, is finite.

We close with a remark concerning the finiteness of the state sum. In order to prove
finiteness of the state sum, one would need the explicit evaluation of the constants $C^{\rho,N,\epsilon}_{kk'}$ in \Ref{Kbound} as functions of the representation labels (see appendix \ref{lorentz}). However, we leave this for further investigation.

\section*{Acknowledgements}

The authors thank Carlo Rovelli and Abhay Ashtekar for helpful remarks on a prior draft,
and a referee for valuable comments, including pointing out reference \cite{fl2004}.
J.E. was supported in part by the Alexander von Humboldt foundation of Germany and NSF grant
OISE-0601844.

\appendix

\section{Proof of finiteness from the matrix element bound}
\label{proofs}

Throughout this appendix we will use the notion of integrability of what we call
a labelled graph.  Given a graph $\Gamma$, we assign
a principal series representation $(\rho_\ell, N_\ell)$ to each link $\ell$, and
to each pair $(n, \ell)$ of a node and incident link, we assign an $SU(2)$ spin $k_{n \ell}$
and a half-integer $m_{n \ell} \in \{-k_{n \ell}, -k_{n\ell}+1, \dots, k_{n\ell}\}$.
The graph $\Gamma$, together with the labels $\rho_\ell, N_\ell, k_{n \ell}, m_{n \ell}$
which we collectively denote by $\Xi$,
is what we call a `labelled graph.'
Given such a labelled graph $(\Gamma, \Xi)$,
choose an arbitrary node $t_*$ in $\Gamma$, and number the nodes in $\Gamma$, starting with $t_*$,
$1, \dots, M$ for convenience. As in section \ref{maintextproof}, denote
links by the unordered pair $(ij)$ of numbers corresponding to the nodes at either end,
and let ordered pairs $ij$ of adjacent nodes denote the choice of a node $i$ and link $(ij)$
incident on it. Then $(\Gamma, \Xi)$ is said to be
an integrable graph if the following quantity is finite:
\begin{equation}
\label{genFdef}
F^\Gamma(\Xi) := \left[\prod_{i=1}^{N} \int_{x_i \in H} \dif x_i\right]
\prod_{\substack{i,j \in \{1, \dots, M\} \\ i<j}}K^{\rho_{(ij)},N_{(ij)}}_{k_{ij}m_{ij},k_{ji}m_{ji}}(x_{i}, x_{j}) ,
\end{equation}
where $K^{\rho_{(ij)},N_{(ij)}}_{k_{ij}m_{ij},k_{ji}m_{ji}}(x_{i}, x_{j})$ is defined as in \Ref{Kdef}.
Note equation \Ref{Fdef} is a special case of equation \Ref{genFdef}
when $\Gamma$ is the boundary of a 4-simplex.

We prove in this appendix the analogues of Lemma 5 and Theorem 2 of \cite{bb}.
Although not all of the analogue of Theorem 3 of \cite{bb} is needed for this paper,
we state it in full as well, for completeness, though without proof, as the proof is an immediate adaptation of that in \cite{bb}\footnote{The parts of the argument of \cite{bb} involving the mathematical details of the ``propagator'' (equation (3) in \cite{bb}) are entirely encapsulated in Lemma 5 and Theorem 2.  In proving analogues of the results of
\cite{bb} for the present case, the only difference is that the relevant propagator is now
equation (\ref{Kdef}) of this paper. This is why only the analogues of Lemma 5 and Theorem 2 need to be fully reproven here.}.

The importance of Lemma 5 is two fold. In the first place, it is important in the proof of Theorem 2, which states that the tetrahedron graph is integrable.  Secondly, and more importantly, it guarantees that, given an integrable graph, every other graph constructed from it by adding a node with at least three legs will also be integrable. This is the first part of Theorem 3. These two conclusions then imply that the 4-simplex graph is integrable, which we state as a corollary. Notice the full content of Theorem 3 in fact proves integrability for a much larger class of graphs. The integrability of these more general graphs may be useful, e.g., for defining versions of the new spin-foam models in which polyhedra more general than 4-simplices are allowed.

\setcounter{lemma}{4} 
\begin{lemma}
If $n\geq 3$, the integral
\begin{equation}
J:=\int_H \; \dif x\; \left|K^{\rho_1 ,N_1}_{k_1k'_1,m_1m'_1}(x,x_1)\right| ... \left|K^{\rho_n ,N_n}_{k_nk'_n,m_nm'_n}(x,x_n)\right| \nonumber
\end{equation}
converges and for any $0<\epsilon <1/3$ there exists $C^\epsilon(\{\rho_i,N_i,k_i,k'_i\})$, $i=1...n$, function of the representation labels, such that for any $(x_1,...,x_n)$,
\begin{equation}
J\leq C^\epsilon(\{\rho_i,N_i,k_i,k'_i\}) \exp\left(-\frac{n-2-n\epsilon}{n(n-1)}\;\sum_{i<j}r_{ij}\right), \nonumber
\end{equation}
where $r_{ij}:=d(x_i,x_j)$.
\end{lemma}
{\startproof
First, using \Ref{Kbound} one has:
\begin{equation}
\left|K^{\rho_1 ,N_1}_{k_1k'_1,m_1m'_1}(x,x_1)\right| ... \left|K^{\rho_n ,N_n}_{k_nk'_n,m_nm'_n}(x,x_n)\right|\leq \left(\prod_{i=1}^n\; C^{\rho_i,N_i,\epsilon}_{k_i k'_i}\right) \; e^{-(1-\epsilon)\sum r_i},
\end{equation}
where $r_i:=d(x,x_i)$. Define $\tilde{C}:=\prod_{i=1}^n\; C^{\rho_i,N_i,\epsilon}_{k_i k'_i}$, then one has
\begin{equation}
J\leq 4\pi\tilde{C}\int_0^\infty \sinh^2 r \dif r e^{-(1-\epsilon)\sum r_i},
\end{equation}
where $r$ is defined as the distance of $x$ from the barycentre of the points $(x_1,...x_n)$. The fact that it exists is object of Lemma 4 in \cite{bb}. From the same lemma, one has
\begin{equation}
\sum r_i\geq nr.
\end{equation}
In addition, defining
\begin{equation}
M:=\frac{1}{n} min_x \sum_i r_i(x),
\end{equation}
one has
\begin{equation}
\sum r_i\geq nM .
\end{equation}
Both inequalities can be used to prove the following bound for J:
\begin{equation}
J\leq 4\pi\tilde{C}C'\;e^{-(n-2-n\epsilon)M},
\end{equation}
for some positive constant $C'$ depending only on $\epsilon$ and $n$.
From the triangle inequality, one has
\begin{equation}
\sum r_i \geq \frac{1}{n-1}\sum_{i<j} r_{ij},
\end{equation}
and
\begin{equation}
M\geq \frac{1}{n(n-1)}\sum_{i<j} r_{ij},
\end{equation}
which then implies the lemma with $C=4\pi\tilde{C}C'$.
\finishproof}

\setcounter{theorem}{1} 
\begin{theorem}
The tetrahedron graph, with any labelling, is integrable.
\end{theorem}
{\startproof
We will show that the following quantity (for any fixed $x_1\in H$ and independent of it) is finite:
\begin{eqnarray}
I & := &\int_{H^3}\; \dif x_2\dif x_3\dif x_4\; \left| K^{\chi_{12}}(x_1,x_2)K^{\chi_{13}}(x_1,x_3)K^{\chi_{14}}(x_1,x_4)\right. \nonumber \\
& & \left.K^{\chi_{23}}(x_2,x_3)K^{\chi_{24}}(x_2,x_4)K^{\chi_{34}}(x_3,x_4)\right| ,
\end{eqnarray}
where $\chi_{ij}$ denotes, for short, the set of labels
$(\rho_{(ij)}, N_{(ij)}, k_{ij}, m_{ij}, k_{ji}, m_{ji})$.
Start by integrating over $x_4$ using Lemma 5,
\begin{equation}
I\leq C^\epsilon(\{\chi_{ij}\})\;\int_{H^2}\;\dif x_2 \dif x_3\; e^{-\frac{1}{6}(1-3\epsilon)(r_{12}+r_{13}+r_{23})}\left|K^{\chi_{12}}(x_1,x_2)K^{\chi_{13}}(x_1,x_3)K^{\chi_{23}}(x_2,x_3)\right|,
\end{equation}
where $r_{ij}=d(x_i,x_j)$. Next, we integrate over $x_3$. Consider the quantity
\begin{equation}
L:=\int_H\; \dif x_3\;  e^{-\frac{1}{6}(1-3\epsilon)(r_{13}+r_{23})}\left|K^{\chi_{13}}(x_1,x_3)K^{\chi_{23}}(x_2,x_3)\right|.
\end{equation}
By \Ref{Kbound}, one has
\begin{equation}
L\leq \tilde{C} \int_H\; \dif x_3\; e^{-(r_{13}+r_{23})(\frac{7}{6}-\frac{3\epsilon}{2})}
\end{equation}
Now, introduce the new coordinate system $(k,l,\phi)$, where:
\begin{equation}
k=\frac{1}{2}(r_{13}+r_{23})\;\; ;\;\; l=\frac{1}{2}(r_{13}-r_{23}),
\end{equation}
and $\phi$ is the angle between the plane containing $x_1,x_2,x_3$ and a given plane containing $x_1$ and $x_2$. Their ranges are: $k\in [\frac{r_{12}}{2},\infty)\,$, $l\in [-\frac{r_{12}}{2},\frac{r_{12}}{2}]$ and $\phi\in [0,2\pi)$. The measure $\dif x_3$ on $H$ in this coordinate system reads (see appendix of \cite{bb}):
\begin{equation}
\dif x_3=2\;\frac{\sinh r_{13} \sinh r_{23}}{\sinh r_{12}}\; \dif k\;\dif l\;\dif\phi .
\end{equation}
In terms of these new coordinates, we have:
\begin{eqnarray}
L&\leq& \frac{\tilde{C}}{\sinh r_{12}}\;\int_0^{2\pi} \dif\phi\int_{-\frac{r_{12}}{2}}^{\frac{r_{12}}{2}}\dif l\int_{\frac{r_{12}}{2}}^{\infty}\dif k\; e^{-k(\frac{1}{3}-3\epsilon)}\nonumber \\
&\leq& \frac{2\pi r_{12} \tilde{C}}{\sinh r_{12}}\; e^{-r_{12}(\frac{1}{6}-\frac{3\epsilon}{2})}
\end{eqnarray}
for $\epsilon <1/9<1/3$. Plugging this in the evaluation of $I$, we get:
\begin{eqnarray}
I&\leq& C' \int\dif r \; r\sinh r \; e^{-r(\frac{4}{3}-3\epsilon)}\nonumber \\
&\leq& C' \int\dif r \; r \; e^{-r(\frac{1}{3}-3\epsilon)},
\end{eqnarray}
which is finite for $0<\epsilon<1/9$ and some constant $C'$ depending on the representation labels $\{\chi_{ij}\}$.
\finishproof}

\begin{theorem}
A graph obtained from an integrable graph by connecting
an extra vertex to the existing labeled graph by at least three
edges, with arbitrary labeling, is integrable.
A graph obtained from an integrable graph by adding extra edges,
with arbitrary labeling, is integrable.
A graph constructed by joining two disjoint
integrable graphs at a vertex is integrable.
\end{theorem}
Using the analogue of Lemma 5 above, the first assertion
follows using the same arguments as in \cite{bb}.
The second and third assertions follow using the same
arguments as in \cite{bb}.

\begin{corollary}
The 4-simplex graph, with any labelling, is integrable.
\end{corollary}

\section{Useful facts about the Lorentz group}
\label{lorentz}

Let $V\in SL(2,\mathbb{C})$, then one has the following decomposition:
\begin{equation}
V=R\; d(r)\; R'\; ,
\end{equation}
where $R,R'\in SU(2)$ and
\begin{equation}
d(r)=B_z(r)=\left(\begin{array}{cc} e^{r/2} & 0  \\
                               0 &  e^{-r/2} \end{array}\right) .
\end{equation}
The Haar measure in this decomposition reads
\begin{equation}
\dif V = \frac{1}{4\pi} \; \dif R \dif R' \sinh^2 r \dif r,
\end{equation}

We complete this appendix with some explicit formulas for the matrices $d^{\rho,N}_{kk'm}(r)$, referred to in the main text. In particular, we show that the asymptotic behavior \Ref{dasymp} holds. We follow closely section (4-5) of \cite{ruhl}. We start with the following useful expression:
\begin{equation}
\label{dmatrix}
d^{\rho,N}_{kk'm}(r)=\left\{\cdots\right\}^{\frac{1}{2}}\;\left(2\sinh r\right)^{-k-k'}\;\sum_{\nu ,\mu}\; c_{\nu\mu}\; e^{\mu r}\;\frac{\sinh\left(r(i\rho/2+\nu)\right)}{(i\rho/2+\nu)\sinh r}\; ,
\end{equation}
where $\nu+k'$ and $\mu+k'$ are integers, and
\begin{equation}
\left\{\cdots\right\}=\left\{(2k+1)(2k'+1)\times\frac{\left(k+\frac{N}{2}\right)!\left(k-\frac{N}{2}\right)!\left(k'+\frac{N}{2}\right)!\left(k'-\frac{N}{2}\right)!}
{(k+m)!(k-m)!(k'+m)!(k'-m)!}\right\} .
\end{equation}
To define the coefficients $c_{\nu\mu}$, it is useful to redefine the summation labels $(\nu,\mu)\rightarrow(a,b)$, while introducing a new sum over integers $(n_1,n_2)$:
\begin{eqnarray}
2b&=&\mu+\nu+k'-k+\frac{N}{2}+2n_1-m \\
2a&=&\nu-\mu+k'+k+m-\frac{N}{2}-2n_1 .
\end{eqnarray}
The sum over $(\nu,\mu)$ can then be traded by a sum over $(n_1,n_2,a,b)$:
\begin{eqnarray}
&&\sum_{\nu\mu}\;c_{\nu\mu}\;(\cdots)=\sum_{n_1,n_2}\; \left(\begin{array}{c} k+m \\ n_1\end{array}\right)\left(\begin{array}{c} k-m \\ n_1-m+\frac{N}{2}\end{array}\right)\left(\begin{array}{c} k'+m \\ n_2\end{array}\right)\left(\begin{array}{c} k'-m \\ n_2-m+\frac{N}{2}\end{array}\right)\nonumber \\
&&\times\sum_{a,b}\;(-1)^{a+b+m-\frac{N}{2}}
\left(\begin{array}{c} k+k'-n_1-n_2+m -\frac{N}{2} \\ a\end{array}\right)\left(\begin{array}{c} n_1+n_2-m+\frac{N}{2}\\ b\end{array}\right)\; (\cdots)
\end{eqnarray}
where all summations extend over the domain where the binomial coefficients do not vanish. From eq. \Ref{dmatrix}, one sees that the asymptotic behavior for $r\rightarrow\infty$ is of the form:
\begin{equation}
d^{\rho,N}_{kk'm}(r)\sim e^{-r(k+k'+1-(\mu+\nu)_{max})},
\end{equation}
for $(\mu+\nu)$ taking its maximal value. One can check that this maximal value is given by:
\begin{equation}
(\mu+\nu)_{max}=k+k'-\left|m+\frac{N}{2}\right|
\end{equation}
which then gives the asymptotic behavior
\begin{equation}
d^{\rho,N}_{kk'm}(r)\sim e^{-r(1+\left|m+\frac{N}{2}\right|)}
\end{equation}
as advertised in the main text.
A last step, which is not necessary for the proof of finiteness but should be very useful for the finiteness analysis of the state sum model, is the evaluation of the maximum of $e^{(1-\epsilon)r} d^{\rho,N}_{kk'm}(r)$, as a function of $r$. This would allow for the explicit expression of the coefficients $C^{\rho,N,\epsilon}_{kk'}$ in \Ref{Kbound} in terms of the representation labels.

\section{Consideration of full triangulation and the gauge-fixing interpretation of the regularization}
\label{gfapp}

In the main text, for brevity, we did not derive the spin foam sum from a
discrete path integral on the full triangulation.
Due to the new nature of the derivation --- specifically the use of non-gauge invariant tensors ---
the derivation of (\ref{Z}) using the full triangulation has a small difference from
the standard derivation.  We review this difference.  With the derivation based on the full triangulation in mind,
we then review the internal gauge-fixing procedure in \cite{fl2004}, which is the standard procedure in lattice gauge theories \cite{creutz}. We will see that this gauge-fixing procedure
cannot be used in our case, but must be modified; the modified procedure will be equivalent to
the regularization proposed in the main text.

Given the parallel transports $V_{tv}$ around the link of a face $f$,
let $U_f(t):= U_f(t,t)$ denote their composition in clockwise order starting at $t$.
The discrete action \cite{eprlett, eprpap, elpr} is
\begin{equation}
\label{discaction}
S_{\text{disc}} = \frac{1}{16 \pi G} \sum_{f \in \Delta}
\tr\left(\left(B_f(t)
+ \frac{1}{\gamma}\dual B_f(t)\right) \Lambda[U_f(t)] \right)
= \sum_{f \in \Delta}
\tr\left(J_f(t)\Lambda[U_f(t)] \right)
\end{equation}
where $\Lambda: SL(2,\C) \rightarrow SO(3,1)$ denotes the
standard 2-1 homomorphism.
Next, for any four $SL(2,\C)$ irreps $(\rho_4, N_4), \dots, (\rho_4, N_4)$, we have the
following resolution of the identity on
\mbox{$\Hil_{N_1,\rho_1} \otimes \cdots \otimes \Hil_{N_4, \rho_4}$}:
\begin{eqnarray}
\ident^{(k_1,m_1) \dots (k_4,m_4)}{}_{(k_1', m_1') \dots (k_4',m_4')} \hspace{-0.5em} &=& \hspace{-0.5em}
\delta^{k_1}_{k'_1} \cdots \delta^{k_4}_{k'_4}
\sum_{w \in \scrB_{\tilde{k}_1,\dots \tilde{k}_4} \hspace{-0.5em}}
w^{m_1 \dots m_4} (w^\dagger)_{m_1' \dots m_4'} \\
\nonumber
&\hspace{-1em}=& \hspace{-1em}\sum_{\tilde{k}_1,\dots \tilde{k}_4} \sum_{w \in \scrB_{\tilde{k}_1,\dots \tilde{k}_4} \hspace{-1em}}
\left[\left(\otimes_{i=1}^4 P_{\tilde{k}_i}\right) \otimes w \right]^{m_1 \dots m_4}
\left[\left(\otimes_{i=1}^4 P_{\tilde{k}_i}\right) \otimes w \right]^\dagger{}_{m_1'\dots m_4'}
\end{eqnarray}
where $\scrB_{\tilde{k}_1, \dots \tilde{k}_4}$ is a fixed orthonormal basis of $\Hil_{\tilde{k}_1} \otimes \cdots \otimes \Hil_{\tilde{k}_4}$
%
%
for each 4-tuple of $SU(2)$ spins $\tilde{k}_1, \dots, \tilde{k}_4$,
and the $P_k$ are as in equation (\ref{projspindef}).
We then compute the partition function for (\ref{discaction})
using the same strategy as in \cite{baez}, except using the resolution
of the identity on the full tensor space, instead of just on the intertwiner space.  This yields
\begin{equation}
\label{ZBF}
\scrZ := \int \prod_{f} \dif J_f(t)
\prod_{(t,v)} \dif V_{tv}
e^{i S_{\text{disc}}[J,V]}
= \sum_{k_{ft}, w_t}
\left[\prod_{f} \sum_{n=0}^{\infty} \int_{-\infty}^\infty \hspace{-1em}(N_f^2 + \rho_f^2)\dif \rho \right]
\prod_{v}
A_v\left[\rho_f, N_f; k_{tf}, w_t\right]
\end{equation}
%
%
where for each vertex $v$, the integrations over the five connection variables $\{V_{vt}\}_{t\in v}$ are
absorbed into the expression for $A_v$ in equation (\ref{bfvertex}).

Now, a discrete gauge transformation is specified by a group element $G_v \in SL(2,\C)$ at each
vertex and a group element $G_t \in SL(2,\C)$ at each tetrahedron, with action
$V_{vt} \mapsto G_v V_{vt} G_t^{-1}$, $B_f(t) \mapsto G_t B_f(t) G_t^{-1}$.
Let us review the gauge-fixing procedure of \cite{fl2004}: we will then see why
the gauge-fixing procedure of \cite{fl2004} does not work in our case, and then we will present
a different procedure.

First, if one does not impose the simplicity constraints, (\ref{ZBF}) is a partition function for
BF theory, so that the gauge-fixing strategy of \cite{fl2004} applies.  One first chooses a maximal tree $T$ of the
1-skeleton of the cell complex $\Delta^*$ dual to the triangulation $\Delta$. Each 1-cell of $T$ is an edge
$e$ dual to a tetrahedron $t$, with parallel transport $V_{vt}V_{tv'}$.
As $T$ contains no closed loops, one can use the aforementioned gauge freedom to fix to the identity
all $V_{vt}$ in $T$.  This is the gauge fixing procedure of \cite{fl2004} (adapted to the present variables).

\begin{figure}\label{gfpic}
\begin{center}\includegraphics[height=1.3in]{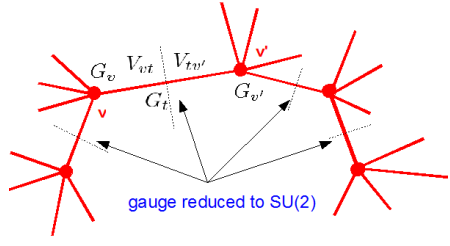}\end{center}
\caption{The figure stands for a general dual triangulation. The gauge invariance is reduced at the tetrahedra.}
\end{figure}

In the new models, however, we must impose the simplicity
constraints (\ref{finalsimp}), $k_{ft} = \frac{N_f}{2} = \frac{\rho_f}{2\gamma}$.
As noted in the main text, because $k_{ft}$ is the quantum number of a non-Lorentz invariant
quantity, these constraints break $SL(2,\C)$ gauge symmetry at the tetrahedra, reducing the gauge there
to $SU(2) \ni G_t \equiv g_t$ (see fig. 1).
As a consequence one will no longer be able to fix completely the group elements on a maximal tree (one may fix the rotation part of the group leaving it a pure boost, but this doesn't help us in the proof of finiteness). At the end of the day, we are able to fix to the identity only one $V_{tv}$ per 4-simplex, which is equivalent to the regularization procedure presented earlier in this paper. Let $V_{t_v v}$ denote the group element in $v$ that we gauge-fix to
the identity.  This gauge-fixing condition implies
$G_v g_t^{-1} = \ident$, i.e. $G_v = g_t$.
Thus the $G_v$ gauge freedom is precisely the gauge that has been fixed by
$V_{t_v v} = \ident$, leaving the $SU(2)$ gauge transformations at the
tetrahedra free.

\end{document}